\documentclass[11pt,twoside,letter]{article}
\usepackage{amsthm, amsmath, amssymb, amsfonts}
\usepackage{mathtools}
\usepackage{booktabs}

\usepackage{bbm}
\usepackage{mathrsfs}
\usepackage{graphicx}

\usepackage{epstopdf}
\usepackage[shortlabels]{enumitem}
\usepackage[colorlinks=true, pdfstartview=FitV, linkcolor=blue, citecolor=blue, urlcolor=blue]{hyperref}
\usepackage[usenames]{color}
\usepackage{url}
\makeatletter\def\url@leostyle{%
	\@ifundefined{selectfont}{\def\UrlFont{\sf}}{\def\UrlFont{\scriptsize\ttfamily}}} \makeatother
\usepackage[margin=1.0in, letterpaper]{geometry}

\newtheorem{theorem}{Theorem}
\newtheorem{proposition}[theorem]{Proposition}

\theoremstyle{definition}
\newtheorem{definition}[theorem]{Definition}

\theoremstyle{remark}
\newtheorem{remark}[theorem]{Remark}

\numberwithin{equation}{section}
\numberwithin{theorem}{section}
\definecolor{Red}{rgb}{0.9,0,0.0}
\definecolor{Blue}{rgb}{0,0.0,1.0}
\def\cL{\mathcal{L}}

\def\bE{\mathbb{E}}
\def\bF{\mathbb{F}}

\def\bP{\mathbb{P}}

\def\bR{\mathbb{R}}

\def\sF{\mathscr{F}}

\def\sS{\mathscr{S}}

\def\bfR{\mathbf{R}}

\def\bfW{\mathbf{W}}

\def\bfb{\mathbf{b}}

\def\bfw{\mathbf{w}}

\newcommand{\1}{\mathbbm{1}}            % preferable way of writing indicator function
\usepackage{tikz}
\usepackage{multirow}
\usepackage{subfig}
\usepackage{caption}

\def\vpi{\boldsymbol{\pi}}

\def\rq{\mathtt{q}}
\title{A groundwater market model}

\makeatletter
\def\and{%
\end{tabular}%
\begin{tabular}[t]{c}}%
\def\@fnsymbol#1{\ensuremath{\ifcase#1\or a\or b\or c\or
		d\or e\or f\or g\or h\or i\else\@ctrerr\fi}}
\makeatother

\author{
	Igor Cialenco\,\thanks{Department of Applied Mathematics, Illinois Institute of Technology
		\newline \hspace*{1.45em}  10 W 32nd Str, Building RE, Room 220, Chicago, IL 60616, USA
		\newline \hspace*{1.45em}  Emails: \url{cialenco@iit.edu}, URL: \url{http://cialenco.com}
		\vspace{0.5em}} \quad and  
	\and
	Michael Ludkovski,\thanks{Department of Statistics and Applied Probability, University of California, Santa Barbara \newline  \hspace*{1.45em}
		Building 479, UC Santa Barbara, Santa Barbara, CA 93106-3110, USA \newline  \hspace*{1.45em}
		Email: \url{ludkovski@pstat.ucsb.edu}, URL: \url{http://ludkovski.pstat.ucsb.edu/}
	}
}

\date{ {\small 
		First Circulated and this version: January 23, 2025\\
}}

\begin{document}

	\maketitle

	\vspace{-2em}
	
	%	\smallskip
	
	{\footnotesize
		\begin{tabular}{l@{} p{350pt}}
			\hline \\[-.2em]
			\textsc{Abstract}: \ & 			
		We introduce the problem of groundwater trading, capturing the emergent groundwater market setups among stakeholders in a given groundwater basin. The agents optimize their production, taking into account their available water rights, the requisite water consumption, and the opportunity to trade water among themselves. We study the resulting Nash equilibrium, providing a full characterization of the 1-period setting and initial results about the features of the multi-period game driven by the ability of agents to bank their water rights in order to smooth out the intertemporal shocks. 			
			
			\\[0.5em]
			\textsc{Keywords:} \ &  groundwater market, groundwater price, Nash equilibrium, water rights, water banking, Pareto optimality.   \\
			\textsc{MSC2020:} \ &  91B76, 91A10, 91B72, 91B70  \\[1em]
			\hline
		\end{tabular}
		
	}

\section{Introduction}\label{sec:intro}
Spurred by the problem of groundwater overextraction and the looming danger of catastrophic aquifer depletion, jurisdictions are enacting legislation to manage large-scale groundwater pumping. For example, the California Sustainable Groundwater Management Act (SGMA) \cite{sgma2014} of 2014 calls for the creation of Groundwater Sustainability Areas (GSAs) to manage groundwater in each basin (roughly corresponding to a local watershed). Each GSA is responsible for adopting a Groundwater Sustainability Plan that quantifies and oversees groundwater usage across the GSA and is collectively governed by the local stakeholders, primarily farmers. In order to align the economic incentives with the respective need to curb existing groundwater overuse, many GSAs are experimenting with groundwater markets, that allow these stakeholders to trade their pumping rights among themselves. Facilitating  water rights trading could play a critical role in transitioning to a sustainable water supply system and supporting mandates such as the SGMA \cite{AyresEtAl2021}. Note that these are not financial markets, but physical exchanges with tight regulations and fixed number of participants, who optimize their revenues but are devoid of speculative motives. 

In this paper we propose a novel stochastic model to study these emerging markets. We adopt a top-down approach to model the groundwater replenishment process, a key feature that captures system-level dynamics, and we formulate 
a  multiperiod stochastic game among economic agents (farmers), in which the agents control the type and quantity of the crops to produce, and how much water rights to buy or sell, contingent on meeting physical and regulatory  constraints. Each agent strives to  maximize a risk-reward functional of her revenue process and the research goal is to study the \emph{equilibrium water rights prices} as an output of a stochastic game without a central planner. From a dynamic perspective, agents face stochastic shocks due to random water allocations (based on precipitation-driven groundwater recharge each year) mitigated by their ability to bank water rights across time, leading to complex non-zero-sum game effects.  With this short notice, we aim to lay the foundation for this important  applied problem, which gives rise to several nontrivial game-theoretic and computational  challenges that we plan to address in the sequel.  We refer to \cite{DinarHogarth2015} for a review of  game  theoretical approaches in water resource management, mostly in a 1-period setup. Closer to our approach are \cite{GanjiEtAl2007a,GanjiEtAl2007} that study optimal reservoir operation policies using stochastic dynamic Nash game with perfect information. % among water users.

\subsection{The Model}
On a filtered probability space $(\Omega, \sF, \bP, \bF= (\sF_t)_{t=0}^T)$, assume that $J$ economic agents, also refereed as farmers, trade within one basin groundwater rights among themselves. Throughout, all processes are assumed $\bF-$adapted. 
We use boldface to denote vectors, e.g.~$\boldsymbol{\varphi}=(\varphi_1,\ldots,\varphi_J)$. Fraktur letters  denote  sums of quantities across agents, e.g. $\mathfrak{W}=\sum_{j=1}^JW_j$.  

Each agent $j$ has the opportunity to produce $K$ types of goods (e.g.~grow different crops), and can sell good $k$ at time $t$ for a net profit of $\overline{f}^k_j(t,\varphi_j^k)$ dollars per $\varphi_j^k$ units of good $j$. We assume that the function $\overline{f}^k_j(t,\cdot)$ are continuous.  Agents also can trade water among themselves at the beginning of each time period $t$. 
Denote by $\psi_j(t)$ the amount of water traded, in acre-feet (ac-ft),  by agent $j$, with  convention that  $\psi_j>0$ means selling water, and $\psi_j<0$ buying water.  Let $p(t)$ be the price of traded water (in \$/ac-ft) at time  $t$. Then, the $j$-th agent P\&L is given by 
\begin{align}\label{eq:L}
L_j(t) := %\underbrace{
\sum_{k=1}^K \overline{f}^k_j(t, \varphi_j^k) \cdot \varphi^k_j(t) + \psi_j(t) \cdot p(t)
\end{align}

The agents control the type and the quantity of the goods to produce and the amount of water to trade, and hence 
$\pi_j(t):= (\boldsymbol{\varphi}_j(t), \psi_j(t)) = (\varphi^1_j(t), \ldots, \varphi^K_j(t), \psi_j(t)),$ are 
the $j$-th agent controls, subject to additional  production constraints 
\begin{align}
	0 \leq n^k_j(t) \leq  & \ \varphi^k_j(t) \leq N_j^k(t) \qquad \text{ for all } k,t.   \label{eq:constr_Bounds} 
\end{align}
The lower bound  $n_j^k(t)$  %serves as a contractual obligation of delivering corresponding goods, or
represents agricultural constraints. For example, a farmer growing perennial crops (walnut orchard) cannot skip a year of production, while a grower of annual vegetables may plant something else instead, or to fallow land altogether.  The processes $ \mathbf{n}(t), \mathbf{N}(t)$ are taken as inputs to the model, and could be deterministic or stochastic. 
The production of one unit of good $k$ requires $a^k(t)$ ac-ft of water. We denote agent's  $j$ water needs for producing the goods at time $t$  by 
$C_j(t) :=\sum_{k=1}^K \varphi^k_j(t) a^k(t).$ 

Let $W_j(t)$ be the total amount of water available to agent $j$ at time $t$. 
The process $\{W_j(t)\}$ is random, with the dynamics modeled below,  and  includes the water rights given by the regulator to agent $j$ at time $t$, as well as the unused water  from the previous periods, but excludes the  water needed to  produce $\varphi_j^k(t)$ goods or the amount of traded water $\psi_j(t)$.  

We assume that  agents cannot use more water  than total available to them for that period, namely, imposing the water budget constraints
\begin{equation}
    -\sum_{i\neq j} W_i(t)  \leq C_j(t) + \psi_j(t) \leq  W_j(t). \label{eq:waterBudgetConstr}
\end{equation}
The agents have the opportunity for groundwater banking, that is, $C_j(t)+\psi_j(t)< W_j(t)$ yielding intertemporal shift in water consumption. Moreover, the traded water amounts must balance out via the market clearing condition 
\begin{equation}\label{eq:mrktClearPsi}
\sum_{j=1}^J \psi_j(t) = 0. 
\end{equation}

 Denote by $\sS_j$ the set of all feasible controls $\pi_j$ of agent $j$, i.e.~stochastic processes satisfying constraints \eqref{eq:mrktClearPsi}-\eqref{eq:waterBudgetConstr}.
Each agent maximizes a risk-reward functional $U_j(\cdot)$ of her revenue, and solves the control problem 
$
\sup_{\pi_j\in\sS_j} \bE \Big[\sum_{s=t}^T U_j\big( L_j(s)\big) \ \big| \,  \sF_t \Big]. 
$
The functions $U_j$  account for each agent's idiosyncratic utility, as well as the temporal discount factor, and are required to be monotone increasing and concave.

\subsection*{Water Budgets Model}
Groundwater available for pumping is driven by the height $H(t)$ of the water table, measured in ac-ft. We develop a tractable dynamic model for $H(t)$, assuming that 
the agents $j=1,\ldots,J$ are the only ones tapping from  this water basin, and equating it for simplicity with available water $H(t) = \mathfrak{W}(t)$, for all $t$.  
The evolution of $H(t)$ is based on the groundwater recharge process $\{R(t)\}$, which is the amount of water (primarily from surface precipitation percolating down) entering the aquifer from time $t$ to $t+1$. We assume that the process $R(t)$  is given exogenously, and follows a Markovian structure.  
A companion study will detail data analyses to justify and calibrate such models for $\{ R(t) \}$ based on publicly available data. The dynamics of the water table height is given by 
$H(t+1) = H(t) + R(t+1) - \mathfrak{C}(t), \quad t=0,\ldots,T-1,  
$ with initial condition $H(0)$. 

We postulate that the regulator  allocates at each period $t$ a fixed proportion
$\theta_j$  of the recharged amount $R(t)$  to agent $j$, based on grandfathered historical rights,  
for some  fixed $\theta_j>0$, such that $\sum_j\theta_j = 1$.  Thus the dynamics of available water for agent $j$ is 
\begin{equation}\label{eq:dynWj}
W_j(t+1) = W_j(t) + \theta_jR(t+1) - C_j(t) - \psi_j(t), \quad t=0,\ldots, T-1, 
\end{equation}
with $W_j(0)= \theta_j H(0)$. Available water increases through new rights tied to the exogenous process $R(\cdot)$ and decreases due to consumption and trading.

Considering from now on only Markovian policies $\pi_j\in\sS_j$, the process $(\boldsymbol{W}(t), R(t))$  is Markov, and  the agents solve  
\begin{equation}\label{eq:Vjw}
% V_j(t, \boldsymbol{w}, r) =
\sup_{\pi_j\in\sS_j} \bE \Big[\sum_{s=t}^T U_j\big( L_j(s)\big) \ \big| \  \boldsymbol{W}(t)=\boldsymbol{w}, \ R(t)=r \Big].
\end{equation}

\section{Equilibrium water rights price}\label{sec:waterprice}
Traditional to non-cooperative games, we assume that the agents play a stochastic dynamic Nash game \cite{BasarOlsder1999} without a central planner, with individual preferences revealed through the maximization problems \eqref{eq:Vjw}.
The fair water rights price process $p^*(\cdot)$ is determined from the Nash equilibrium  point $(\vpi^*, p^*) = ( (\varphi_j^*(t), \psi_j^*(t)), (p^*(t))$, $j=1,\ldots, J, \ t=0,\ldots,T$,
such that no agent can gain by deviating, assuming the other agents keep their strategies unchanged. 
Formally, given the feasible  strategies  $\vpi_{-j}$ chosen by agents other than $j$, agent $j$  
maximizes her expected payoff 
\[
A_j(\pi_j, \pi_{-j},p) = \bE \Big[\sum_{s=0}^T U_j\big( L_j(s; \pi_j(s),p(s))\big)\Big],
\]
with respect to ${\pi}_j\in\sS_j$. Additionally, we  introduce a fictitious player, called the \emph{price-setter}, whose expected payoff is 
$
A_{J+1}(p, \vpi) := \bE \big[ \sum_{s=0}^{T} p(s)  \sum_{j=1}^J \psi_j(s)\big],
$
and who is choosing only $p$. The feasible set $\sS_j$, and hence the expected payoff of farmer $j$, depends on $\vpi_{-j}$. In particular, due to the market clearing conditions, the strategies of the other farmers determine how much water farmer $j$ is trading, $\psi_j = -\sum_{i\neq j} \psi_i$.  Often, such Nash games are called generalized Nash games. 
\begin{definition}
The pair $(\vpi^*, p^*)$  is called a Nash equilibrium (NE) if 
\begin{equation}\label{eq:Nash2}
\begin{split}
 \forall j=1,\ldots, J, \quad \forall \pi_j' \in \sS_j, \quad
&A_j(\pi_j^*, \vpi_{-j}^*,p^*) \geq A_j(\pi_j', \vpi_{-j}^*,p^*), \\
\forall v' \in \sS_{J+1}, \quad & A_{J+1}(p^*, \vpi^*) \geq A_{J+1}(v', \vpi^*).
\end{split}
\end{equation}
\end{definition}
Next we note  that given the additive structure of the P\&L \eqref{eq:L}, it is enough to solve the optimization problem \eqref{eq:Vjw} in terms of the water consumption amounts $C_j$'s and the amounts traded $\psi_j,\ j=1,\ldots,J$. 
Given a water budget $C$ denote by $G_j(t,C)$ the farmer's $j$ maximum profit from producing the goods subject to her production constraints, namely 
\begin{align}\label{eq:G}
G_j(t,C):=\max_{\boldsymbol{\varphi}_j} 
\sum_{k=1}^K \overline{f}^k_j(t, \varphi_j^k) \cdot \varphi^k_j
\end{align}
subject to \eqref{eq:constr_Bounds} and $\sum_k a^k(t)\varphi_j^k=C$. If $f_j^k$ are time homogeneous, then we simply write $G_j(C)$. Note that $G_j(C)$ is well defined for all $C$ such that
\begin{equation}\label{eq:BoundsC}
\underline{c}_j(t):=\sum_k a^k(t) n_j^k(t) \leq C \leq \sum_k a^k(t) N_j^k(t) =: \bar{c}_j(t).   
\end{equation}
Assuming that there are no production bounds, i.e. $n_j^k=0, \ N_j^k=+\infty$, we  make a reasonable assumption that $G_j$ is monotone increasing  on $[0,M]$, and continuously differentiable on $(0,M)$ for some sufficiently large $M$.  Moreover, under \eqref{eq:constr_Bounds}, $G_j(C)$ is monotone and continuous on $(\underline{c}_j,\overline{c}_j)$ 
Generally speaking, the functions $G_j$ do not have an explicit form, but they can be computed offline, and from computational point of view we can effectively assume that $G_j$ is given. 
With slight abuse of notation, we re-write \eqref{eq:L} as $L_j(t) = G_j(t,C_j) + \psi_j(t) p(t)$ and denote the strategy of agent $j$ by $\pi_j(t)=(C_j(t), \psi_j(t))$.  Then, clearly the feasible controls $\pi_j\in\sS_j$ are stochastic processes satisfying  \eqref{eq:mrktClearPsi}, \eqref{eq:waterBudgetConstr} and \eqref{eq:BoundsC}, and the control problem \eqref{eq:Vjw} remaining the same.

Taking self-sufficient farmers as the economic baseline,  trading is a ``bonus''. Namely, if the water price $p$ is not ``satisfactory'', then farmers can recede into self-consumption based on their allocation. Indeed, the Nash concept implies that if nobody else is interested in trading, then regardless of the wishes of the given farmer, they will have to not trade as well (a side trade being a bilateral deviation that is beyond Nash stability requirements). Thus, assuming 
\begin{equation}\label{eq:minWaterNeed}
\underline{c}_t(t) \leq \theta_j R(t), \quad \forall t,    
\end{equation}
the existence of a NE  equilibrium is apparent. Indeed, for sufficiently large $p(t)$, all agents would like to sell water, and nobody will be willing to buy. In this case, $\psi_j(t)=0, \  \forall j$, and each agent will have to meet her production needs at $t$ solely through her assigned water rights, which is feasible thanks to  \eqref{eq:minWaterNeed}. The latter assumption is not necessary;  it could happen that some agents either deliberately bank water to meet future lower bounds of production  \eqref{eq:constr_Bounds} or reach their upper limits $N_j(t)$ and have no use for their residual rights. But in either case, since nobody is willing to buy water for $p(t)$, no trading occurs. Similarly for $p(t)$ low enough, all agents would like to buy, yielding the similar no-trading outcome as above. 
All these water prices and respective consumption schedules form NE. Having a non-unique NE is a typical situation, and generally speaking there is no unique or canonical method for choosing one.

\subsection{One period market model} In a static model, the NE can be described explicitly, creating a baseline understanding of the structure of NE in general. 

\begin{proposition}\label{prop:NE1} 
Assume that $\underline{c}_j\leq \theta_j R=W_j$ for all $j$. Then, for any $p\geq 0$, there exists $\vpi$, such that $(\vpi, p)$ is a NE.      
\end{proposition}
\begin{proof} 
Note that in a one period setup there is no banking of water, and farmers would like either to sell all excess water, or buy a desired amount. Hence, for a fixed $p$, ignoring \eqref{eq:mrktClearPsi}, each farmer  solves the optimization problem 
\begin{equation}\label{eq:C-p}
\max_{C_j} \left\{ G_j(C_j) + (W_j - C_j)p\right\}, \qquad
\textrm{s.t. \ }     \underline{c}_j \leq C_j \leq \overline{c}_j.  
\end{equation}
Since  $G$ is continuous, \eqref{eq:C-p} has a maximizer, and we denote by $\widetilde{C}_j(p)$ the smallest one.

We note that if $\widetilde{C}_j(p)\leq W_j$, then the farmer $j$ would prefer to sell  up to $W_j-\widetilde{C}_j(p)$ units of water. Given the strategies of other players, she may sell less, say $\psi_j$, in which case the remaining unsold water $W_j-\psi_j$ will be used for production, thanks to $G_j(\cdot)$ being increasing, until reaching the upper production bounds.   
Assuming that $W_j\leq\overline{c}_j$ all available water will be used up. 

If $\widetilde{C}_j(p)\geq W_j$ for all $j$,  then each farmer would like to buy water in addition to her assigned water rights $W_j$, and thus nobody will trade. Hence, optimal strategy is  $\pi_j^* = (W_j, 0) \ \forall j$, and $(\boldsymbol{\pi}^*,p)$ is a NE. By similar arguments, if $\widetilde{C}_j(p)\leq W_j$ for all $j$, then  $\pi_j^* = (W_j, 0)$ is a NE. For all other prices $p$, some agents will be willing to buy and some agents will be willing to sell water, and any $\boldsymbol{\psi}$ satisfying 
\[
\sum_{j:W_j\geq C_j^*(p)} \psi_j \1_{[0, W_j - \widetilde{C}^j(p)]}(\psi_j)   = - \sum_{j: W_j\leq \widetilde{C}_j(p)} \psi_j\1_{[\widetilde{C}_j(p)-W_j,0]}(\psi_j), 
\]
yields a NE. Clearly, such $\psi_j$'s  exist. This completes the proof. 
\end{proof}
\begin{remark}
Generally speaking, it is enough to have $\sum_j \underline{c}_j\leq \sum_j W_j$, for the existence of a NE, but it may happen, that for some $p$, some agents may not be willing to sell, and thus some agents may not meet the production lower bounds. Describing all such prices $p$ is possible, albeit cumbersome. 
Proposition~\ref{prop:NE1} and its proof do not hold true, in general, for a multiperiod market where banking plays a critical role.  
\end{remark}

If in addition we assume that $\sum_j \underline{c}_j < R <  \sum_j \overline{c}_j$, then there exists $p^\circ$ such that 
\[
\sum_j \widetilde{C}_j(p^\circ) = R.
\]
A feasible strategy $(\vpi,p)$ is called Pareto optimal if there is no other feasible strategy $(\vpi',p')$ such that 
\[
\forall j \ A_j(\vpi', p') \geq  A_j(\vpi, p), \quad \textrm{and} \quad \exists i \ 
A_i(\vpi', p') >  A_i(\vpi, p). 
\]
In view of the above, $p^\circ$ is the price at which each agent achieves her maximal profit in the one period setup, and thus the NE $(\widetilde{\boldsymbol{C}}(p^\circ),p^\circ)$ is Pareto optimal. 

Next, we consider a specific, yet general, net profit function $\overline{f}_j^k(t, \varphi_j^k) = f_j^k (\varphi_j^k)^{\alpha_j^k-1} - q_j^k\varphi_j^k$, for some constants $f_j^k, q_j^j\geq 0$ and $\alpha_j^k\in(0,1)$. 
The concave and increasing power function $f_j^k (\varphi_j^k)^{\alpha_j^k-1}$ is meant to capture the market price impact of selling goods,  while the linear term $q_j^k\varphi_j^k$  quantifies the idiosyncratic costs, such as production cost, cost of  pumping  water, etc.  Without loss of generality we consider linear utility functions $U(x)=x$. If there are no production constraints, the Lagrangian for agent $j=1,\ldots,J$, and for a fixed $p$, becomes 
\[
\cL(\boldsymbol{\varphi}_j; \lambda_j) = \sum_k (f_j^k(\varphi_j^k)^{\alpha_j^k} - q_j^k\varphi_j^k )+ (W_j 
- \sum_k a^k \varphi_j^k)p  
+ \lambda_j(R - \sum_{i,k} a^k\varphi_i^k)
\]
and the combined f.o.c.\ conditions for agents $j=1,\ldots,J$,  are 
\begin{align}
    & f_j^k \alpha_j^k (\varphi_j^k)^{\alpha_j^k-1} -q_j^k - a^k p - a^k\lambda_j = 0, \label{eq:FOC3}\\
&  R  - \sum_{i,k} a^k\varphi_i^k = 0, \qquad 
\varphi_j^k\geq 0, \quad \quad k=1,\ldots, K. \label{eq:FOC4}
\end{align}
Let  $d_j^k:=(a^k/\alpha_j^k f_j^k)^{1/(\alpha_j^k-1)}$, and $e_j^k:=q_j^k/a_k$. 
Then, for any $\lambda_j \geq - p - e_j^k$,
\begin{align}
& \varphi_j^k = (p+\lambda_j+e_j^k)^{\frac{1}{\alpha_j^k-1}} d_j^k , \quad k=1,\ldots, K, \label{eq:KKT1}\\
& R =  \sum_{k,j}(p+\lambda_j+ e_j^k)^{\frac{1}{\alpha_j^k-1}} a^k d_j^k. 
\label{eq:KKT2}
\end{align} 
Clearly, for any $p\geq 0$, there exist $\lambda_j\in\bR$, $j=1,\ldots,J$, such that \eqref{eq:KKT2} is true, and hence \eqref{eq:KKT1} is satisfied. Moreover, since the optimization problem at hand is convex with affine constraints, these conditions are  sufficient for $\boldsymbol{\varphi}$ to be the joint maximizer.

For a fixed price $p>0$ announced  by the price setter, assuming  that agent $j$ has no constraints on available water to buy/sell, her optimal production is given by 
\[
\varphi_j^k = (p+e_j^k)^{\frac{1}{\alpha_j^k-1}} d_j^k,
\]
which decreases as function of $p$, i.e.\ lower price will yield larger production, and hence larger water consumption. 
To meet these optimal production rates, agent $j$ would buy water if  
\[
W_j \leq \sum_k a^k \varphi_j^k = \sum_k a^k (p+e_j^k)^{\frac{1}{\alpha_k^j-1}} d_j^k.  
\] 
Due to the monotonicity of the power function, there exists unique price $\widetilde{p}_j$ such that the above inequality becomes equality. Thus, for any $p\leq \underline{p} = \min_j  \widetilde{p}_j$
no agent would be willing to sell water. The agent will consume all the allocated water $W_j$, and by direct calculations we obtain the optimal production rates $\varphi_j^k = (\widetilde{p}_j+e_j^k)^{\frac{1}{\alpha_j^k-1}} d_j^k$. This implies that the agent $j$ chooses the Lagrange multiplier $\lambda_j$ in \eqref{eq:KKT1} corresponding to the NE such that $p+\lambda_j = \widetilde{p}_j$, hence $\lambda_j>0$ making the agent $j$ artificially boost the profit from producing the goods. 

Similarly, if $p> \overline{p} = \max_j \widetilde{p}_j$, each agent prefers to  sell water, thus  no water is traded, and the multipliers in \eqref{eq:KKT1} are chosen so that $\lambda_j<0$, which can be interpreted as the agent $j$ produces goods pretending that the water price is the indifference price. 

From here, it is apparent that to reach the maximum profit, for a fixed price $p$, each agent $j$ should choose  $\lambda_j=0$. Taking $\lambda_j = 0 \ \forall j$, we get the unique price $p^\circ$ that clears the market 
\[
\sum_{j,k} (p^\circ+e_j^k)^{\frac{1}{\alpha_j^k-1}} a^k d_j^k = R. 
\]
Clearly, this price $p^\circ$ is Pareto optimal. We also note that if there were a central planner who would implement the cooperative solution of maximizing the total production revenue of all farmers $\sum_j G_j(C_j)$  subject  to $\sum_j C_j = R$, this yields the same f.o.c.
\[
\varphi_j^k = (\lambda^{SO} + e_j^k)^{\frac{1}{\alpha_j^k-1}}d_j^k, \qquad \sum_{j,k} a^k \varphi_j^k = R,
\]
and hence the same optimal production rates $\varphi^k_j$ corresponding to $p^\circ$. Moreover, we can interpret $p^\circ$ as the shadow cost of water (i.e. the Lagrange multiplier for the budget constraint) from the central planner's perspective. 
 Note that the social optimum is driven by the total available water $\mathfrak{W}$ and agent-specific allocations $W_j$ play no role beyond cash transfers within the ``subsidiary'' farmers in the community.

Similar computations and arguments can be extended to the case with nontrivial production bounds $n_j^k,N_j^k$. For example, assuming that  $\sum_j \underline{c}_j < R <  \sum_j \overline{c}_j$, the social optimum  exists and is computed from KKT conditions
\begin{equation}\label{eq:phi-v*}
\varphi_j^k = n_j^k\vee (\lambda^{SO}+ e_j^k)^{\frac{1}{\alpha_j^k-1}} d_j^k \wedge N_j^k, \qquad \sum_{j,k} a^k \varphi_j^k =R.
\end{equation}
Correspondingly, the max profit functions $C_j(C)$ in \eqref{eq:G} are computed through  $\varphi_j^k  = n_j^k \vee (\lambda_j(C) + e_j^k)^{\frac{1}{\alpha^k_j-1}} d_j^k  \wedge N_j^k$, where $\lambda_j(C)$ are the Lagrange multipliers, such that $\sum_{j,k} a^k \varphi_j^k =R$.

\subsection{Illustrative Example and Comparative Statics}\label{sec:one-period}

To illustrate the above arguments, we present a numerical case study featuring two producers $j=1,2$ and two goods $k=1,2$, see Table \ref{tab:params}. We work in a static one-period setting with linear utility $U(x)=x$.

\begin{table}
    \centering
{\small 
\begin{tabular}{r|c|c|c|c|c|c}
   &  $\alpha_j^\cdot$ & $f_j^\cdot$ & $q_j^\cdot$ &  $a_j^\cdot$  & $n_j^\cdot$ & $N_j^\cdot$\\ \hline
  Farmer 1 & (0.75, 0.8) & (7, 10) & (2, 4) & (1, 2) & (5, 5) & (40, 30)\\
  Farmer 2 & (0.75, 0.9) & (9, 9) & (2, 4) & (1, 2)  &  (5, 5)&  (40, 30) \\ \hline
\end{tabular}
}
   \caption{Parameter values for the numerical examples in Section 2 and 3.}
    \label{tab:params}
\end{table}

\begin{figure}
    \includegraphics[width=0.48\textwidth,trim=0.1in 0.25in 0.1in 0in]{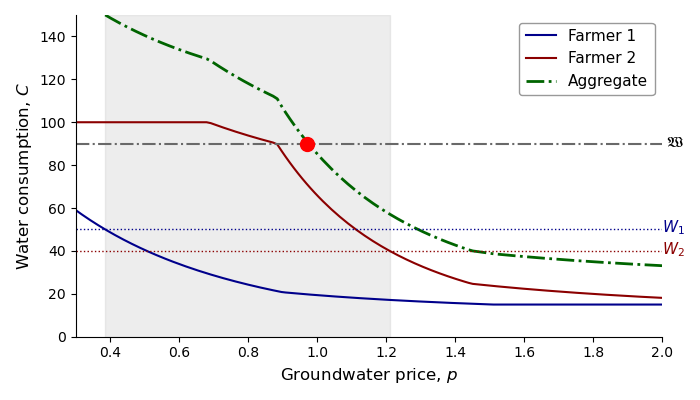} $ \quad$
    \includegraphics[width=0.48\textwidth,trim=0.1in 0.25in 0.1in 0in]{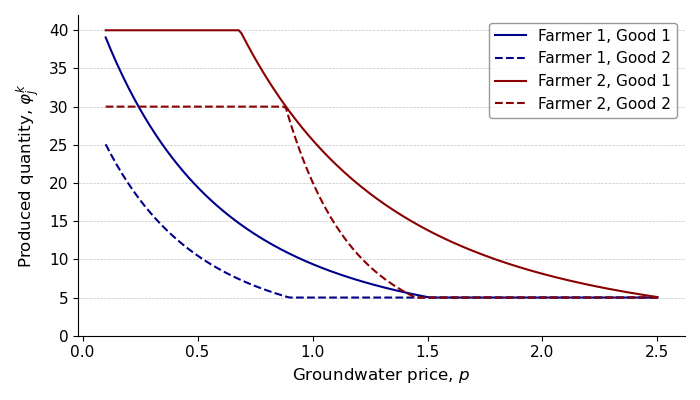}
    \caption{One-period groundwater market with two agents and two goods. \emph{Left:}   groundwater consumption amounts $C_j(p)$ as a function of water price $p$. Shaded area indicates the prices for which trades occur. Efficient price $p^\circ=0.975$ corresponds to setting $\mathfrak{C}(p^\circ) = \mathfrak{W}$ (horizontal gray line) total available water. Horizontal dotted lines indicate farmer individual allocations $W_1, W_2$. \emph{Right:} optimal production quantities $\varphi_j^{k,*}(p)$ given groundwater market price $p \in [0.1, 2.5]$.
\label{fig:one-period-phi}}
\end{figure}

Maximizing \eqref{eq:C-p}, we obtain the optimal  consumption curves $C_j(p)$ for any given water price $p$. The left panel of Figure~\ref{fig:one-period-phi} shows these water consumption curves $C_j(p)$  desired by each agent given the set groundwater price $p$, as well as the resulting aggregate consumption $\mathfrak{C}$. The maps $p \mapsto C_j(p)$ are monotone, and are of piecewise-power type in this example.  The Pareto optimal $p^\circ$ emerges from the water budget constraint. Namely, $p^\circ$ is characterized as the unique crossing point of the aggregate consumption curve $\mathfrak{C}(p)$ with the level $\mathfrak{W}$,  with the respective trading amounts backed out according to $\psi_j = W_j-C_j(p^\circ)$. 

One may also directly tackle the optimization of the production quantities $\varphi_j^k$, bypassing the aggregate revenue $G_j$, which yields $\varphi_j^{k,*}(p)$ as in \eqref{eq:phi-v*}, with $\lambda^{SO}=p$. 
The right panel of Figure \ref{fig:one-period-phi} illustrates how the above optimal production levels $\varphi_j^{k,*}(p)$ vary as a function of $p$; being a power function this dependence is strictly monotone. Moreover, we observe how production constraints simply clip the respective production levels above or below, see e.g., the role of the upper bound $N_2^1=40$ for $p \le 0.7$. We then have $C_j(p) = \sum_k a^k_j \phi^{k,*}_j(p)$ that are plotted on the left. 

On the right of Figure~\ref{fig:one-period-phi}  we see that $p^\circ=0.975$ at which price Farmer 1 is selling ($C_1(p^\circ) < W_1$) and Farmer 2 is buying, $C_2(p^\circ)> W_2$. More generally,
trading occurs for $p \in [0.385,1.210]$; cf.~shaded area in Figure~\ref{fig:one-period-phi}. For $p$ below $\underline{p}=0.385$, both agents wish to buy water beyond what was allocated to them, for $p$ above $\bar{p}=1.210$ both agents would rather sell water than consume it.

Water allocations are the primary driver of agent profits and water trades. As noted above, the Pareto optimal (PO) groundwater price $p^\circ$ is a function of the total water  $\mathfrak{W} = \sum_j W_j$ only and does not depend on individual allocations $W_j$.  Hence, the impact of changing $W_j$ propagates through two channels: change in $p^\circ$ due to change in $\mathfrak{W}$; change in $\psi_j$'s. Indeed, more water for agent $j$ increases her profit (as she is able to produce/trade additional amounts), but also impacts the other farmer. Namely, larger supply of water $\mathfrak{W}$ will necessarily lower the PO $p^\circ$ and therefore benefit farmers who are buying and conversely hurt those who are selling $\psi_j < 0$.

The dependence of $p^\circ$ on $\mathfrak{W}$ can be understood as the intersection of the aggregate water consumption $\mathfrak{C}(p)$ with the level $y=\mathfrak{W}$, see right panel of Figure~\ref{fig:one-period-phi}. Since $\mathfrak{C}(p)$ decreases in $p$, $\mathfrak{C}'(p^\circ)^{-1}$ gives the rate of change of $p^\circ$ as additional allocation is procured. The resulting impact on the agents is asymmetric: first, lower prices lead to higher consumption and therefore higher profit $G_j(C(p^\circ(\mathfrak{W})))$. Second, lower prices reduce the trading revenue. The third and final effect is the direct impact of increased $W_j$ on farmer $j$: additional allocation increases proportionally her revenue by $p^\circ$ (understood as getting marginally more of a good that is trading for $p^\circ$). While the first and last effects are unambiguous, the second one depends on whether agent $j$ is a buyer or seller. Moreover for agent $i \neq j$, the last effect is moot, and the impact of the price reduction vis-a-vis higher profitability is parameter-specific.

\section{Multi-period Market}\label{sec:multi-period}

We now turn our attention to the multi-period problem \eqref{eq:Vjw}. We consider risk-neutral agents that maximize $\sum_{s=0}^T \bE[ L_j(s)]$.

Compared to the single-period model, $T>0$ creates opportunities for groundwater banking, i.e.~intertemporal shift in consumption. Each farmer is able to consume less water in period $t$ in order to have more water in the future. In turn, banking creates additional competition among the agents. Banking by agent $j$ benefits her (by having additional water available in the future, her expected profits rise) but hurts other agents $i \neq j$ (additional allocation to player $j$ lowers the groundwater price in the future and lowers their future profits), see discussion in the previous section. Consequently, the agents engage in a non-cooperative game. To illustrate the resulting NE, we consider a 2-period model, with $t=0$ (``first'') and $t=1$ (``second'' and last) time points, starting  with initial allocation $\bfW(0)=\bfw$ and a random allocation $\bfW(1)$ in the second period given by \eqref{eq:dynWj}.

To keep track of the banking amounts, we introduce 
\[
b_j(t) = W_j(t) - C_j(t) - \psi_j(t),
\]
such that $0\leq b_j(t)\leq \mathfrak{W}(t)$, 
which represents how much water Farmer~$j$ banks in period $t$ and hence is carried over to the next period.  Then, the dynamics \eqref{eq:dynWj} becomes $W_j(t+1) = R_j(t+1) + b_j(t)$, with $R_j(t+1):=\theta_jR(t+1)$.  To capture the potential value of banking, we view all quantities as a function of the water allocation $\bfR=(R_1,\ldots,R_J)$. Hence, we re-write, $p^*(t, \bfR_t), \varphi_j^k(\bfR_t), L_j(\bfR_t)$, etc., treating $b_j$ as an additional control.

\begin{remark} In practice, banking is allowed 
only for a finite number of periods, creating memory effects during agents' decisions. This issue is moot in the present 2-period model where the only banking is from $t=0$ to $t=1$. 
\end{remark}

We focus on the  special case where we postulate that all prices are based on matching supply and demand, i.e.~are of Pareto type. Denote by $V_j(t, \bfw)$ the one-period profit of agent $j$ under the PO equilibrium and given the time-$t$ allocation vector $\bfw$. 
In that case, for any period-1 allocation $\bfW(1)$, we have the unique $p^\circ(1,\bfW(1))$ described in Section \ref{sec:one-period}, as well as the corresponding game payoffs $V_j(1,\bfW(1))$ for each farmer. Since future equilibria are now uniquely characterized, the multi-period problem may be solved backward via dynamic programming. 

Banking allows each farmer to trade-off between having more water today or more water next period. Hence, equilibrium in the banking strategies corresponds to optimizing that trade-off simultaneously for all farmers. To make the presentation concrete, we consider $J=2$ farmers. 
In that case Nash equilibrium is easiest to approach 
via
a \emph{fixed point} $\{b^*_j(0), j=1,\ldots, J\}$ such that $b^*_j(0)$ is the best response given $b^*_{-j}(0)$. We note that this is a non-zero-sum general game; hence no structural results are available beyond existence of a NE in mixed strategies \cite{BasarOlsder1999}. For the case of two farmers, finding a fixed point can be reduced to examining the best response curves $\bar{b}_2 \mapsto B_1( \bar{b}_2)$ and $\bar{b}_1 \mapsto B_2( \bar{b}_1)$ and looking for their crossing points. 

\begin{figure}[!htb] \centering
\vspace*{-0.25in}
\begin{tabular}{cc}
\begin{minipage}{0.43\textwidth}
    \includegraphics[width=0.97\textwidth,trim=0.5in 0.95in 0.5in 0.4in]{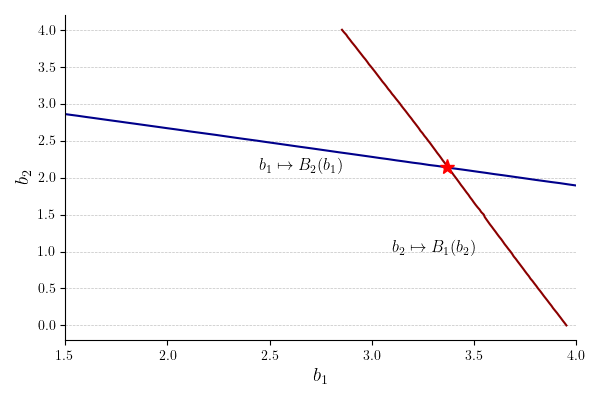}
\end{minipage}
& \begin{minipage}{0.49\textwidth}
{\footnotesize $$ %\hspace*{-6pt}
\ \  \begin{array}{cc|cccc|c}
 \\ \toprule
    & t=0 & \omega_1 & \omega_2 & \omega_3 & \bE[\cdot]  & A_j\\\midrule
     \multicolumn{7}{c}{\text{No banking}}  \\ \midrule
     V_1 & 68.74 & 49.18 & 62.24 & 70.76 & 60.72 & 129.47 \\
V_2 & 75.85 & 51.04 & 67.11 & 78.64 & 65.60 & 141.45 \\
p^* & 0.97 & 1.29 & 1.06 & 0.95 & 1.10 & -- \\
\midrule
 \multicolumn{7}{c}{\text{w/Banking}}  \\ \midrule
V_1 & 66.38 & 52.45 & 64.78 & 72.95 & 63.39 & 129.77 \\
V_2 & 72.76 & 54.71 & 70.32 & 81.61 & 68.88 & 141.64 \\
p^* & 1.00 & 1.23 & 1.03 & 0.93 & 1.06 & -- \\
\bottomrule
\end{array}
$$} \end{minipage}
\end{tabular}
    \caption{Left: Best response curves $\bar{b}_{-j} \mapsto B_j( \bar{b}_{-j})$
     for the two-period market with 2 farmers. The unique crossing point $(3.367, 2.142)$ corresponds to the equilibrium banking strategies. Right: impact of banking, with the last column listing the equilibrium total profit, $A^*_j = V_j(0) + \bE[ V_j(1) | R(0)]$. \label{fig:two-period}}
\end{figure}

Specifically, given initial allocations $\bfw$, for any $\bar{b}_1,\bar{b}_2 \in [0, \mathfrak{W}]$ define
\begin{align} \label{eq:farmer1-2period}
  B_1(\bar{b}_2):=  &\arg\sup_{b_1 \ge 0} V_1 \left(0, \bfw - \begin{bmatrix} b_1 
    \\ \bar{b}_2 \end{bmatrix} \right) + \bE \left[ V_1 \left(1, \bfR(1) +
    \begin{bmatrix} b_1 \\ \bar{b}_2 \end{bmatrix} \right) \Big| \ R(0) = r\right]; \\ \label{eq:farmer2-2period}
  B_2(\bar{b}_1) :=      &\sup_{b_2 \ge 0} V_2 \left(0, \bfw - \begin{bmatrix} \bar{b}_1 \\ b_2 \end{bmatrix}\right) + \bE \left[ V_2 \left(1, \bfR(1) +
    \begin{bmatrix} \bar{b}_1 \\ b_2 \end{bmatrix} \right) \Big| \ R(0) = r\right] \ .
\end{align} 
Equilibrium $(b_1^*, b_2^*)$ is characterized by $b^*_j = B_j(b_{-j}^*), j=1,2$. The resulting period-0 water price then emerges endogenously $p^*(0) = p^*(0, \bfw - \begin{bmatrix} b^*_1 
    \\b^*_2 \end{bmatrix}) $. Similarly, the water consumptions $C^*_j(0)$ and the trading amounts $\psi^*_j(0)$ are determined along with $p^*(0)$. Since $b_j \in [0, \mathfrak{W}(0)]$, the optimization problems in \eqref{eq:farmer1-2period}-\eqref{eq:farmer2-2period} are of a scalar function on a bounded interval, so a solution always exists. However, equilibrium may not be interior and corner solutions of $b^*_j=0$ or $b^*_j = \mathfrak{W}(0)$ (buy up everything available and bank it all) are possible.  Moreover, as  illustrated in Figure \ref{fig:two-period}, we expect the above best response curves to be monotone, hence there is a unique fixed point.      
    
\begin{remark}
 As is clear from \eqref{eq:farmer1-2period}-\eqref{eq:farmer2-2period}, to determine an equilibrium, each farmer $j$ must anticipate and take into account all possible next-stage outcomes, namely $p^*(1, \bfw_1)$ and her future payoffs $V_j(1,\bfw_1)$'s across \emph{all} potential $\bfw_1$. Hence, the role of the regulator is essential, as they can steer, via pre-announcing the full map $p^*(1, \bfw_1)$, the farmers' behavior in period 0. Above we focus on the most non-intrusive regulator that lets the market match supply and demand, i.e.~does not interfere in the market forces under all circumstances, $p^*(1, \bfw_1) \equiv p^\circ(\bfw_1)$. However, a spectrum of other regulator strategies are possible. For example, the regulator could pre-commit to shutting down trading (through a very high $p(1)$)
if $R(1)$ turns out to be low (but otherwise ``not interfere''), which would force agents to bank today to be able to survive on their own in the second period if need be.
\end{remark}

In Figure~\ref{fig:two-period},  we
assume that the period-1 recharge consists of $M=3$ possible states
$$R(1) = \left\{ \begin{aligned}
    50 & & \text{with prob. } \rq_1= 1/9;\\
     75 & & \text{with prob. }\rq_2=4/9;\\
      95 & & \text{with prob. } \rq_3=4/9,
\end{aligned}  \right. \qquad\qquad \text{while}\quad \mathfrak{W}(0) = 90.$$
The allocations for the agents are specified by $\theta_1(t) = 0.6, \theta_2(t) = 0.4 \ \forall t$.
Note that in the second period $\bE[R(1)] < \mathfrak{W}(0)= 90$, so that there is on average
more scarcity (and hence higher market price), incentivizing farmers to postpone their consumption from period-0.

To solve \eqref{eq:farmer1-2period} we employ nested optimization:
\begin{enumerate}
    \item Create a subroutine that determines, for any groundwater allocation $\bfw$, the $V_j(1,\bfw)$'s by evaluating the aggregate water demand $\mathfrak{C}(p)$ as a function of $p$ (see Fig.~\ref{fig:one-period-phi}), and then finding through numerical one-dimensional optimization $p^\circ_1(\bfw) = \mathfrak{C}^{-1}(\mathfrak{W})$. 

   \item Create a subroutine to solve for the best response $B_j(\bar{b}_{-j}(0))$ for agent $j$ in period 0, given the other's banking policy $b_{-j}(0)$. 
   The conditional expectation in \eqref{eq:farmer1-2period} reduces to a  weighted sum  $\sum_m \rq_m V_j \left(1, R_j(1)(\omega_m)+[ b_j, \bar{b}_{-j}]^\top
   \right)$ over $m=1,2,\ldots, M$ (number of states of $R(1)$), where for each term we have to determine the corresponding $p^\circ(1; \omega_\ell)$ using the subroutine from  Step 1;

  \item Find the fixed point $(b_1^*, b_2^*)$ and the resulting banking equilibrium groundwater price $p^*(0)$ and consumption amounts $C^*_j(0)$.

\end{enumerate}

\medskip

In the example, we obtain $b_1^* = 3.367$ and $b_2^* = 2.142$. Namely, in period-0, Farmer 1 utilize their allocation of $W_1(0)=54$ to consume $\varphi_1^1+ 2\varphi_1^2 = 19.33$ units, sell $31.30$ units to Farmer~2 at price $p^*(0)=1.004$ and bank the remaining $3.367$ units. In turn, Farmer~2 starts with allocation of $W_2(0)=36$; she consumes $\varphi_2^1+ 2\varphi_2^2 = 65.16$, and banks $b_2(0)=2.142$, thanks to buying $31.30 = 65.16+2.14-36$ units from Farmer 1. In period-1, the potential allocations would be $(b_1^*, b_2^*)+\{ (30,20), (45,30), (57,38)\}$. Observe that without banking, Farmer 1 on average has $\bE[ R_1(1)] = 48.67$ in period-1 compared to starting out with $W_1(0)=54$; after banking she has $W_1(0)-b_1^* = 50.63$ initially and $\bE[R_1(1)]+b_1^* = 52.03$ in period-1. Farmer 2 similarly smoothes out to $\bE[R_2(1)]+b_2^*=34.59$ vis-a-vis $W_2(0)-b_2^*=33.86$. 

\subsection{Role of trading and banking}

To provide insights into the 2-period market, we consider the two sub-cases of: (a) farmers cannot trade but may bank and (b) farmers cannot bank but may trade.  Inability to bank decouples the periods temporally, 
leading to a sequence of 1-period games already studied above.
In the former case, we have a system of decoupled one-agent problems based on \eqref{eq:G} for each agent, who optimize their period-0 banking amount $\beta_j$ (with the hard period-0 water budget constraint of $C_j(0) = w_j(0)- \beta_j$) to maximize total expected profits 
\begin{align} \label{eq:1agent-2period}
    \sup_{\beta_j \in [0, w_j(0)]} \left\{ G_j(0,w_j(0)- \beta_j) + \bE \left[ G_j(1, \theta_j R(1) + \beta_j) \, |\ R(0)=r \right] \right\}.
\end{align}
We obtain $\beta_1 = 3.180, \beta_2 = 2.504$. Thus, compared to the competitive equilibrium $(b_1^*, b_2^*)$, when left to themselves Farmer 1 would bank less and Farmer 2 would bank more. 

\begin{remark}
The setting of \eqref{eq:1agent-2period} also arises when, for instance, the regulator announces very large $p$ for both periods (say a constant $p$ that is independent of $\bfw(1), R(1)$). In that case, the outcome is no trading, forcing agents to self-optimize their intertemporal consumption.
\end{remark}

The right panel of Fig.~\ref{fig:two-period} compares the  evolution of the equilibrium market price $p^*(t; \cdot)$ and farmer profits $V_j(t; \cdot)$ when banking is allowed or not. We observe that both Farmers benefit from banking and that banking increases prices in period-0 and reduces them in period-1, in particular mitigating the potential price spike in the ``drought'' scenario $\omega_1$.

Notably, due to the ambiguous impact of $w_{-j}$ on $V_j$, it is possible that banking can hurt farmers, since it allows their competitors to ``hoard'' water rights and hence impact (negatively) their future profitability. As a result, a rational behavior by Farmer 1 to smooth out her profits across time may simultaneously negatively impact other farmers who were better off with the original temporal allocations. 

\section{Conclusions}\label{sec:conclusions}

We have introduced the groundwater market model, setting the stage for further analysis of this societally important nascent policy problem. Numerous directions need to be pursued next. Theoretically, it is necessary to study different potential types of dynamic equilibria and their implications, including working with open-loop and closed-loop price-setters, the role of dynamic programming versus direct {open-loop optimization over the vectorized extended-form strategies across multiple periods (i.e.~over the strategy $\pi_j(0), \pi_j(1; \omega_m), m=1,\ldots,M$ for the 2-period setting}).
Numerically, new algorithms are necessary to treat multi-period settings. In a companion follow-up work we will be presenting a machine-learning inspired approach for the latter, which replaces the brute force approach of Section \ref{sec:multi-period} with a more scalable scheme. Namely, one can avoid nested optimization by training a statistical surrogate (for example a neural network) for the mappings $(b_1, \ldots, b_J) \mapsto \bE[ V_j(t+1,  \bfR(t+1) + \bfb) | R(t) = r] $ for each $j$, as a function of banked amounts. On a related note, an algorithm is needed to find fixed points for banking amounts for $J>2$ agents, for example through alternately optimizing for $\check{b}^{(\ell+1)}_j(0)=B_j(\check{b}^{(\ell)}_{-j}(0)), \ell=1,\ldots$, while cycling through $j=1,\ldots, J$.

\section*{Acknowledgments}
IC  acknowledges support from the National Science Foundation grant DMS-2407549, and ML acknowledges support from the National Science Foundation grant DMS-2407550.

\bibliographystyle{alpha}
%\bibliography{WaterAllocation-24-07-01}
\newcommand{\etalchar}[1]{$^{#1}$}

\end{document}